\documentclass[runningheads]{llncs}
\usepackage[T1]{fontenc}
\usepackage{graphicx}
\usepackage{hyperref}
\usepackage{float}
\usepackage{booktabs} 
\usepackage{wrapfig}
\usepackage{multirow}
\usepackage{pifont}

\usepackage{amsmath,amsfonts,amssymb,dsfont}
\usepackage{url}
\usepackage{todonotes}
\usepackage{subfig}
\usepackage[authoryear,longnamesfirst]{natbib}

\usepackage{mathtools}
\usepackage{tabularx}
\usepackage{algpseudocode}
\usepackage{algorithm}
\algnewcommand\algorithmicforeach{\textbf{for each}}
\algdef{S}[FOR]{ForEach}[1]{\algorithmicforeach\ #1\ \algorithmicdo}

\usepackage{url}

\usepackage[dvipsnames]{xcolor}

\usepackage{multicol}

\usepackage[table]{xcolor}
\def\R{\mathbb{R}}

\def\N{\mathcal{N}}
\def\m{\mathrm{m}}

\def\m{{\bf m}}

\def\rr{{\bf r}}

\def\our{MeshSplats}

\usepackage{color}

\begin{document}
\title{\our{}: Mesh-Based Rendering with Gaussian Splatting
Initialization}
%
%
\author{
Rafał Tobiasz
\inst{1, 3}\orcidID{0009-0002-9265-6148} \and
Grzegorz Wilczyński
\inst{1,3}\orcidID{0009-0002-6053-4410} \and
Marcin Mazur\inst{1}\orcidID{0000-0002-3440-8173} \and
Sławomir Tadeja\inst{2}\orcidID{0000-0003-0455-4062} 
\and
Weronika Smolak-Dyżewska\inst{1}\orcidID{0009-0009-1454-3157} \and
Przemysław Spurek\inst{1,3}\orcidID{0000-0003-0097-5521}
}
\authorrunning{Rafał Tobiasz et al.}
%
\institute{Faculty of Mathematics and Computer Science, Jagiellonian University in Kraków, Poland \and
Department of Engineering, University of Cambridge, United Kingdom\and
IDEAS Research Institute, Poland
\\
}
\maketitle              

\vspace{-0.7cm}
\begin{figure*}
\centering
\includegraphics[width=\linewidth]{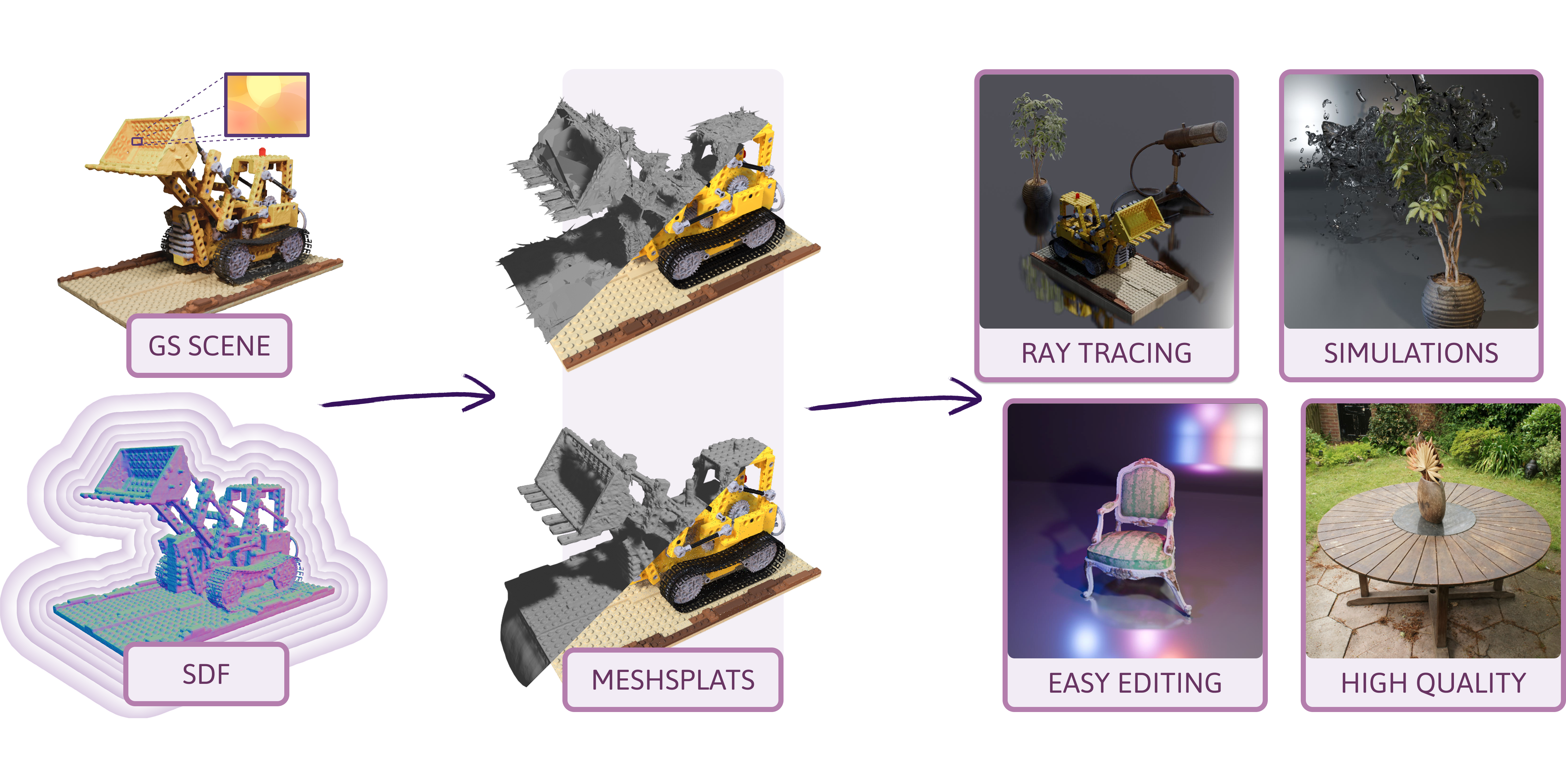}
\caption{
An overview of \our{} pipeline. GS scene (optionally mesh extracted via SDF method) is converted to \our{}. With additional optimization, this representation facilitates advanced downstream applications such as ray tracing, physics simulations, interactive editing, and high-quality rendering.
}
\label{fig:teaser}
\end{figure*}

\vspace{-0.7cm}

\begin{abstract}
Gaussian Splatting (GS) is a recent and pivotal technique in 3D computer graphics. GS-based algorithms almost always bypass classical methods such as ray tracing, which offer numerous inherent advantages for rendering. For example, ray tracing can handle incoherent rays for advanced lighting effects, including shadows and reflections. To address this limitation, we introduce \our{}, a method which converts GS to a mesh-like format. Following the completion of training, \our{} transforms Gaussian elements into mesh faces, enabling rendering using ray tracing methods with all their associated benefits. Our model can be utilized immediately following transformation, yielding a mesh of slightly reduced reconstruction quality without additional training. Furthermore, we can enhance the quality by applying a dedicated optimization algorithm that operates on mesh faces rather than Gaussian components. Importantly, \our{} acts as a wrapper, converting pre-trained GS models into a ray-traceable format. The efficacy of our method is substantiated by experimental results, underscoring its extensive applications in computer graphics and image processing.
\end{abstract}

\section{Introduction}
\label{submission}

The efficient representation of 3D objects is crucial to computer graphics. Classical techniques employ meshes for convenient storage and rapid rendering \citep{foley1994introduction}. Moreover, different ray tracing methods can be applied to these meshes to handle incoherent rays for secondary lighting effects, such as shadows and reflections \citep{raytracing2019}.


Unfortunately, training mesh-like representations directly on 2D images pres\-ents significant challenges. Gaussian Splatting (GS) framework \citep{kerbl20233d} addresses this problem by representing a 3D scene as a set of Gaussian primitives characterized by parameters such as mean, covariance, opacity, and color, often expressed in terms of spherical harmonics (SH). This representation can be effectively trained to achieve high-quality reconstructions and real-time rendering. However, GS is a rasterization technique that utilizes Gaussian projections instead of ray tracing. This inherent characteristic of GS presents a challenge when attempting to incorporate lighting, shadow, or reflection effects \citep{kerbl20233d,moenne20243d}.
Typically, rasterization has been employed to achieve real-time performance by approximating the formation of images. In contrast, ray tracing has enabled comprehensive, high-quality rendering, but it typically works with mesh-based representations.
A potential approach to address this limitation is to integrate GS training with ray tracing, as proposed in \citep{moenne20243d,byrski2025raysplats,von2025linprim}. Such solutions enable the modeling of light and shadow effects, but require dedicated renderers. 

To solve these problems, we introduce \our{}, a method that directly converts GS into a mesh-like representation that can be rendered in existing computer graphics environments such as Blender\footnote{\url{https://www.blender.org}} or Nvdiffrast\footnote{\url{https://nvlabs.github.io/nvdiffrast}} \citep{munkberg2022extracting}. 
Specifically, \our{} is a wrapper, which transforms Gaussian components into disjoint meshes with colors and opacities. These meshes do not approximate the shape of the objects, but they allow for ray tracing-based rendering with light and shadow effects (see Fig.~\ref{fig:teaser}). 
We can fine-tune such meshes to reduce minor artifacts appearing when ellipses are transformed into polygons or triangles.

\begin{figure*}[t!] 
\label{fig:meshsplats_method}
    \centering
        \includegraphics[width=\linewidth]{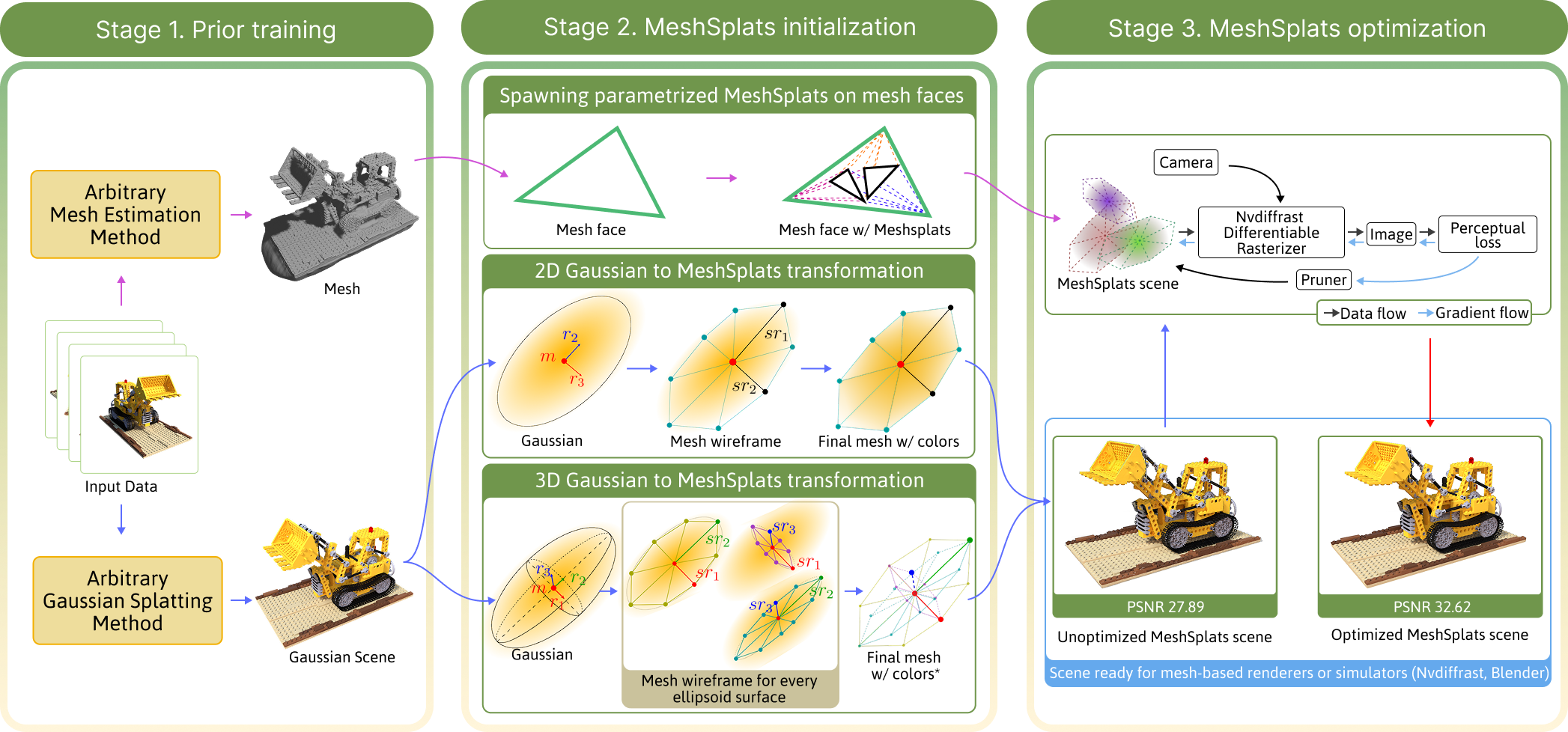}
        \makebox[\columnwidth][c]{%
        } 
        \caption{
        \our{} model operates in two modes. First, with a Gaussian Splatting (GS) prior, we train a GS-based model and convert the resulting ellipses into fan-type meshes, with optional fine-tuning to correct artifacts. Second, using a mesh prior, we train individual triangles directly on a mesh estimated by existing tools \citep{rosu2023permutosdf,wang2021neus}.
        }
    \label{fig:modle}
\vspace{-0.35cm}
\end{figure*}

\our{} works across all 2D and 3D GS-based models (see Fig.~\ref{fig:modle}), but it is particularly effective for flat Gaussians, which exhibit morphologies comparable to meshes. Consequently, converting \our{} representations into mesh faces is more straightforward than with classical 3D Gaussians. Furthermore, we propose an additional method that can incorporate a mesh prior, instead of a GS one, to model object shapes~\citep{rosu2023permutosdf,wang2021neus}. In this setting, \our{} employs two levels of meshes. The first is a classical mesh, estimated by external tools \citep{rosu2023permutosdf,wang2021neus}, dedicated to capturing object geometry. The second, generated by \our{}, consists of disjoint triangle faces with color and opacity attributes, positioned on the geometry mesh (see Fig.~\ref{fig:mesh_modification}).

\begin{figure}\label{fig:msvsms2}
    \centering
        \includegraphics[width=0.7\linewidth]{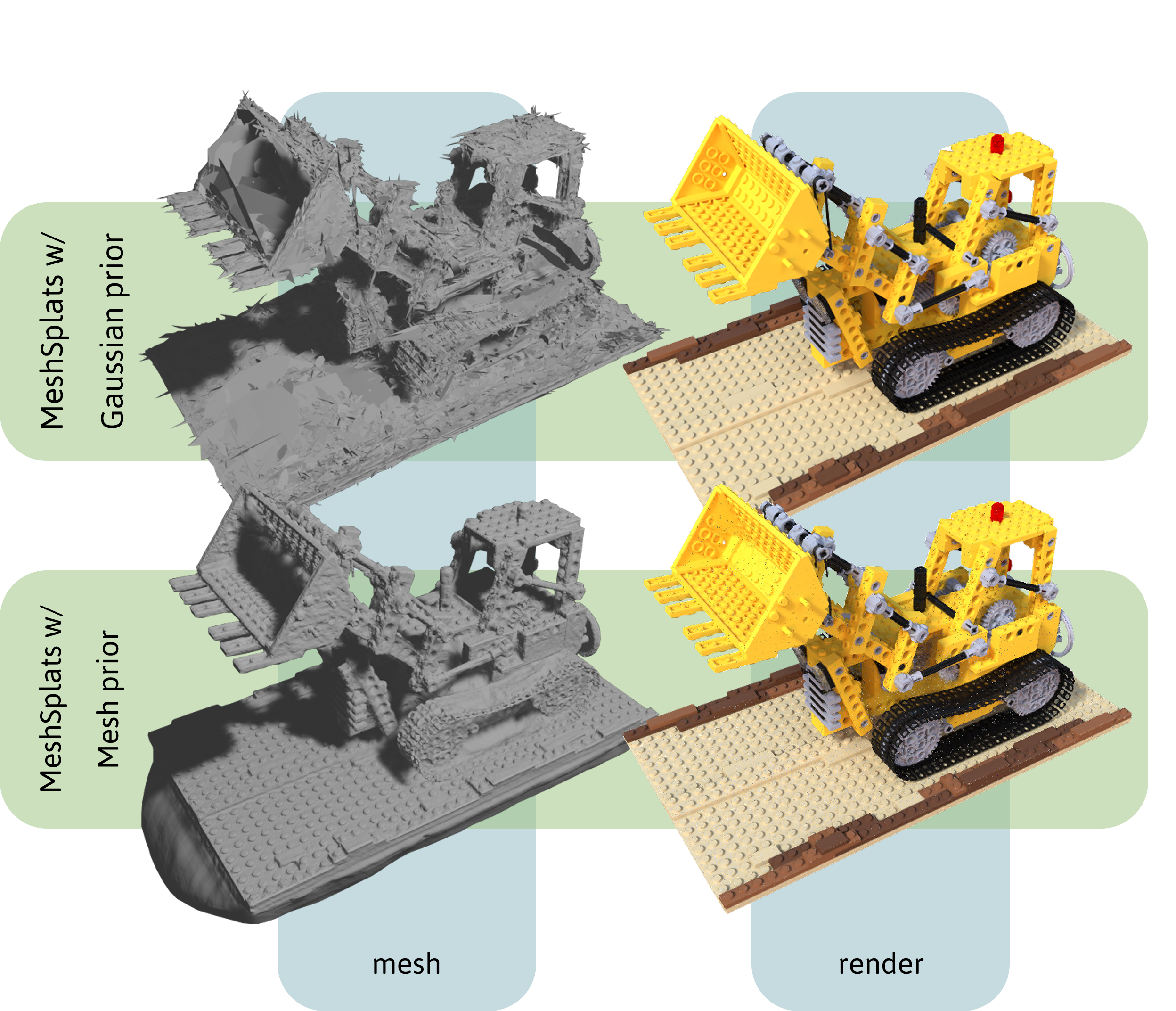}
        \caption{Our model offers two distinct usage scenarios. The first scenario employs a GS prior to wrap-around GS-based models, generating disjoint meshes with minimal tuning. In the second scenario, \our{} incorporates a mesh prior by leveraging meshes estimated from external tools.}
    \label{fig:two}
\vspace{-0.3cm}
\end{figure}

After applying \our{} to the GS scene, the resulting mesh is immediately compatible with standard graphics tools such as Blender and Nvdiffrast, enabling straightforward editing and rendering. Our approach supports two distinct scenarios, see Fig.~\ref{fig:two}. In the first scenario, \our{} leverages a GS prior to act as a wrapper around GS-based models, producing disjoint meshes with minimal tuning. This configuration delivers high-quality renderings but offers limited flexibility for manual editing. In the second scenario, \our{} incorporates a mesh prior derived from external tools. Although this setup requires more extensive fine-tuning, it allows for direct manual modifications to the mesh. Both configurations seamlessly integrate lighting and shadow effects within our framework, as shown in Fig.~\ref{fig:editing}.

In summary, our work contributes the following:
\begin{itemize}
    \item we introduce \our{}, a novel method that facilitates the conversion of GS with flat Gaussians into a mesh-based representation, allowing further integration of dedicated tools;
    \item \our{} is capable of being rendered in traditional rendering environments, thus obviating the necessity of employing specialized GS renderers;
    \item we show that \our{} optimization pipeline effectively reduces artifacts and refines fine geometric details, leading to enhanced geometric accuracy and photorealistic rendering.
\end{itemize}

\section{Related Works}

\paragraph{Gaussian Splatting and Rendering Limitations}
3D Gaussian Splatting (3DGS) \citep{kerbl20233d} was a breakthrough, representing scenes as anisot\-ropic Gaussians optimized by differentiable rasterization. Extensions such as Mip-Splatting \citep{yu2024mip} reduced aliasing, while 2DGS \citep{huang20242d} improved surface alignment by restricting the Gaussians to 2D manifolds. However, these methods rely on spherical harmonics (SH) to approximate view-dependent effects, leading to blurry reflections under complex lighting. Recent work such as GaussianShader \citep{jiang2024gaussianshader} and 3DGS-DR \citep{ye20243d} incorporated environment maps for reflections but were limited to distant lighting. 3iGS \citep{tang20253igs} introduced illumination fields via tensorial factorization, but bounded scene assumptions limit real-world applicability. Crucially, all these methods rely on rasterization, which struggles with the incoherent rays required for advanced effects like shadows and inter-reflections.

\paragraph{Ray Tracing with Gaussian Representations}

Incorporating lighting, shadow, or reflection effects into GS is challenging \citep{kerbl20233d,moenne20243d}, primarily due to fundamental differences between rasterization and ray tracing. While rasterization, commonly used for real-time rendering, relies on image approximation techniques, ray tracing enables high-fidelity rendering but typically operates on mesh-based representations.

Various recent efforts have tried to overcome the limitations of rasterization. 3D Gaussian Ray Tracing (3DGRT) integrates GS training with ray tracing \citep{moenne20243d} and yields excellent visual effects by incorporating light reflection and shadows. It also enables the integration of 3D scenes with mesh-based models. 3DGRT uses bounding primitives to encompass each Gaussian, which facilitates the efficient application of the ray tracing model to these primitives rather than to the Gaussians directly. In RaySplats~\citep{byrski2025raysplats}, the authors use ellipses as an approximation of Gaussians instead of bounding primitives. The above methods use ray tracing approaches for rendering and can model light effects, but require dedicated rendering environments. LinPrim \citep{von2025linprim} employs linear primitives such as octahedra and tetrahedra for differentiable volumetric rendering. This method utilizes mesh-based primitives instead of Gaussians. The model is capable of producing renders of superior quality. However, it necessitates a specialized environment for both training and rendering.  
EnvGS \citep{xie2024envgs} introduces environment Gaussians to model reflections but remains limited to GS-specific rendering pipelines. Similarly, IRGS \citep{gu2024irgs} proposes differentiable 2D Gaussian ray tracing for inverse rendering but requires complex Monte Carlo sampling. These methods highlight the potential of ray tracing and simultaneously inherit the structural limitations of GS, such as dependence on custom renderers and approximations that hinder generalization.


\section{\our{} with Gaussian Splatting Prior}

\begin{figure}[t!] 
    \makebox[\columnwidth][c]{%
        \begin{tabular*}{\dimexpr\columnwidth-1.7cm}{@{\extracolsep{\fill}} c c c }
            \ \ \ \ GT & \ \ \shortstack{\our{}\\w/ Nvdiffrast} & 3DGS
        \end{tabular*}
    }\\[1mm] 
    \centering
    \includegraphics[width=\columnwidth]{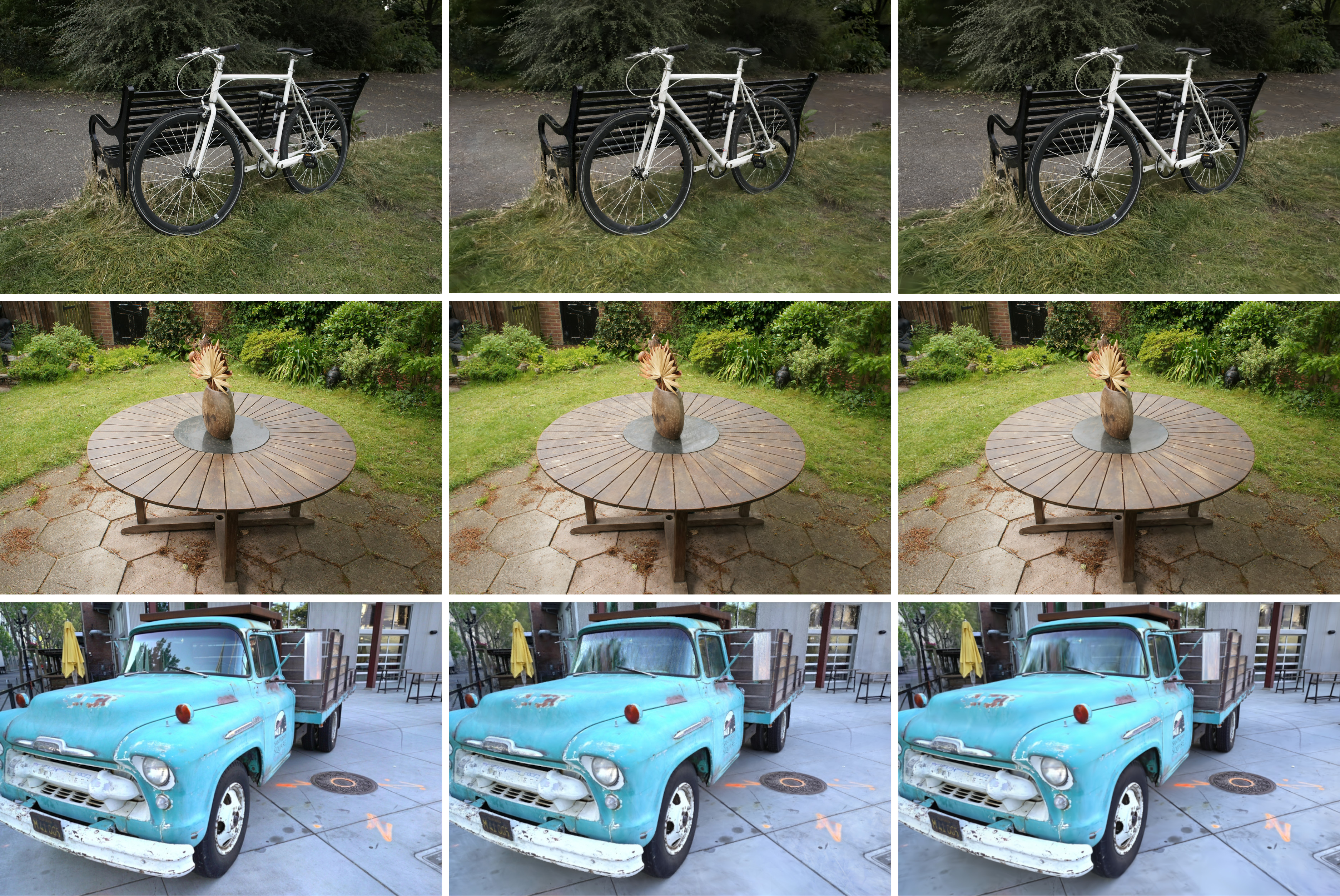}
    \caption{Qualitative comparison: the ground truth, \our{} output, and 3DGS without spherical harmonics. The first two rows use the Mip-NeRF360 dataset, while the last row uses the "Truck" scene from Tanks and Temples, with resolution parameters set to 4, 4, and 1, respectively.
}
    \label{fig:big-scene_nvdiffrast}
\vspace{-0.4cm}
\end{figure}

The core of our approach consists of three straightforward steps. First, we use classical GS to create a collection of Gaussians. Next, we apply \our{} to transform the Gaussian elements into separate mesh faces that approximate the shape of the Gaussians while preserving color and opacity. This step results in a high-quality reconstruction (see Fig.~\ref{fig:big-scene_nvdiffrast}), although minor artifacts may still be visible. Finally, we refine this representation similarly to GS, using separate mesh faces instead of Gaussians. The scheme of the \our{} transformation is shown in Fig.~\ref{fig:modle}.


\paragraph{Gaussian Splatting}

The GS technique constructs a 3D scene using a set of 3D Gaussians, each defined as:
\begin{equation}\label{eq:gs}
(\N(\m,\Sigma), \sigma, c),
\end{equation}
where $\m$ is the mean (position), $\mathbf{\Sigma}$ is the covariance matrix, $\sigma$ is the opacity, and $c$ is the color. Colors are typically represented with Spherical Harmonics (SH)~\citep{fridovich2022plenoxels,muller2022instant}, but can be replaced with standard RGB at a slight cost to view-dependent quality. In \our{}, we use RGB colors for mesh compatibility, enabling external lighting at the expense of minor quality loss.

\paragraph{\our{} Transformation for Flat Gaussians}

\begin{figure}[t!] 
\label{fig:meshsplat2}
    \centering
        \includegraphics[width=0.7\linewidth]{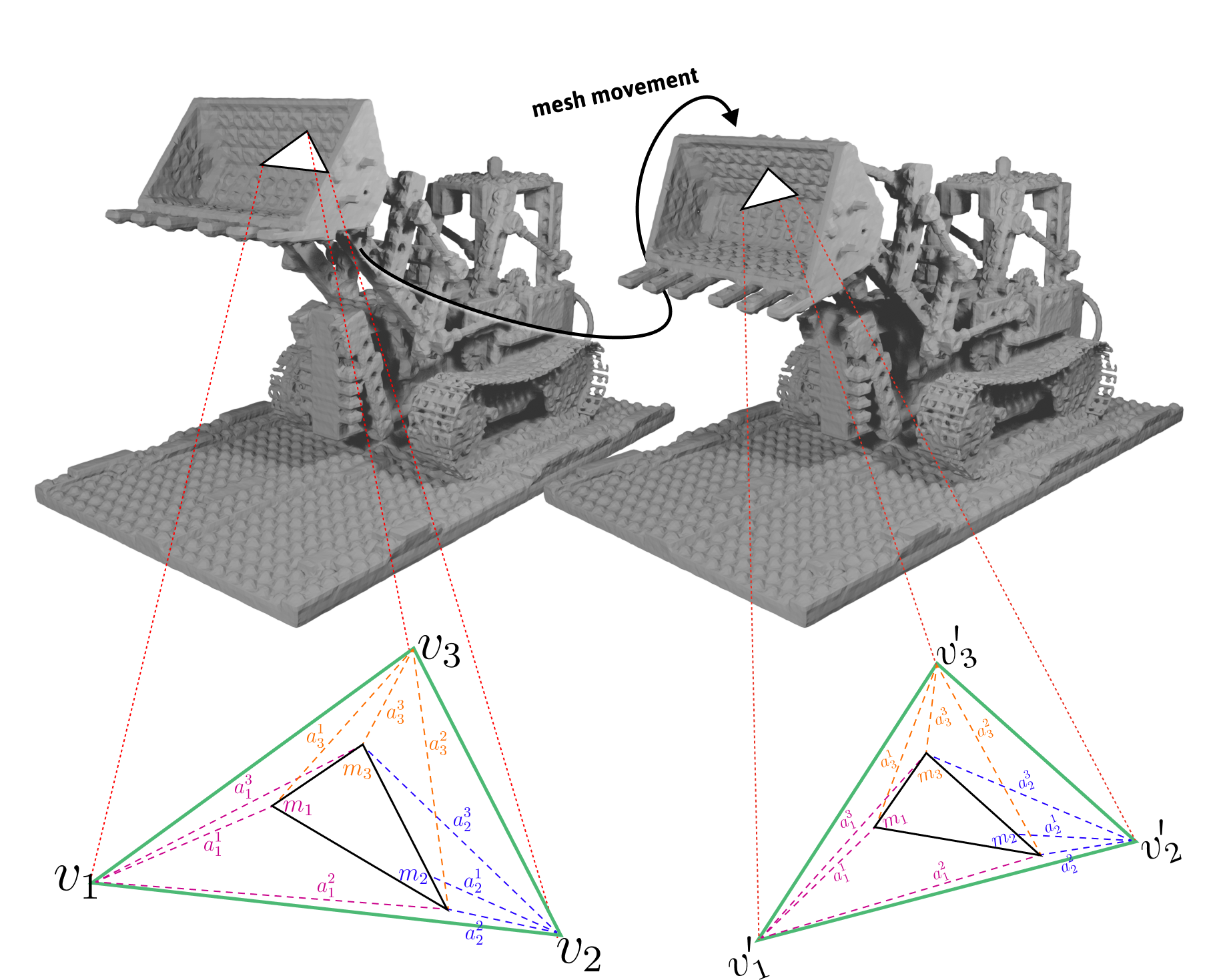}
        \caption{\our{} can operate using a mesh prior estimated by external tools. Here, two levels of meshes are employed: a geometry mesh that models the overall shape of the object, and the second, generated by \our{}, consists of disjoint triangles parameterized by the geometry mesh. This representation enables manual modification.}
    \label{fig:mesh_modification}
\vspace{-0.3cm}
\end{figure}

As mentioned above, in the classical GS model, each element is characterized by a collection of parameters, including a covariance matrix $\Sigma$, which is factorized as
$
\Sigma = RSSR^T,
$
where $R$ is the rotation matrix, and $S$ is a diagonal matrix containing the scaling parameters. There are few approaches that use flat Guassians \citep{guedon2023sugar,waczynska2024games,huang20242d}, but we use the GaMeS representation \citep{waczynska2024games}, given as:
$
\N(\m,R,S),
$
where $\m$ is the mean of the Gaussian, $S=\mathrm{diag}(s_1,s_2,s_3)$, with $s_1=\varepsilon$, and $R$ is the rotation matrix defined as $R=[\rr_1,\rr_2,\rr_3]$, with $\rr_i \in \R^3$. Thus, Gaussian components can be approximated by mesh faces. Our model uses a simple mesh constructed from a polygon inscribed in an ellipse, as shown in Fig.~\ref{fig:modle}.

The transformation of flat Gaussians into meshes is achieved through a series of steps controlled by three hyperparameters: a scale multiplier $scale\_mul$, a number of triangles per Gaussian $no\_triag$, and an opacity multiplier $opac\_mul$. Each Gaussian is represented as a mesh centered at its mean position, with its shape and orientation determined by its scale and rotation parameters. Given that the Gaussians are treated as 2D structures, the scale ($s_i = \epsilon$, where $s_i$ is negligible) and its corresponding rotation vector are discarded, reducing the problem to two dimensions.

First, two orthogonal vectors defining the major and minor axes of an ellipse are computed as follows:
\begin{equation}\label{eq:2d_scaledrot}
\begin{aligned}
    \text{scaled\_rot}_1 &= scale\_mul \cdot \exp(s_2) \cdot \rr_2, \\
    \text{scaled\_rot}_2 &= scale\_mul \cdot \exp(s_3) \cdot \rr_3,
\end{aligned}
\end{equation}
where $\rr_2$ and $\rr_3$ are the second and third columns, respectively, of the rotation matrix $R$, and $scale\_mul = 2.7$. Next, $no\_triag$ points are generated at the boundary of the ellipse and used to construct the mesh. This is done by calculating $no\_triag$ evenly distributed angles $\theta_i$ in the range $[-\pi,\pi]$. For each angle $\theta_i$, the vertex $\text{v}_i$ is computed as follows:
\begin{equation}
    \text{v}_i = \m + \cos(\theta_i) \cdot \text{scaled\_rot}_1 + \sin(\theta_i) \cdot \text{scaled\_rot}_2.
\end{equation}
The mesh is then constructed as a triangle fan, where all triangles share a common origin point at the ellipse's center. Each triangular face is defined by the center point and two consecutive boundary vertices, resulting in $no\_triag$ triangles per Gaussian. This structure approximates the flat Gaussian as a set of connected triangles. In all experiments, we used $no\_triag = 8$.

In our method, color is assigned to each vertex in the mesh, and all vertices inherit the color of the corresponding Gaussian. This ensures that the mesh visually represents the Gaussian's color in the final rendering. Gaussians are trained without SH, guaranteeing compatibility with mesh renderers. On the other hand, opacity is assigned to each vertex in the mesh, with all vertices inheriting the opacity of the corresponding Gaussian. For the boundary vertices, the opacity is scaled by $opac\_mul$ to create a linear interpolation of the opacity across the mesh. This approximates the decrease in Gaussian opacity, with the rate controlled by the $opac\_mul$ parameter. In all experiments, $opac\_mul = 0.2$. 

Through this process, each flat Gaussian is transformed into a mesh that preserves its spatial extent, orientation, and visual properties. The hyperparameters $scale\_mul$, $no\_triag$, and $opac\_mul$ provide control over the fidelity and appearance of the resulting triangle soup, thereby enabling a flexible and accurate spatial representation.

\paragraph{\our{} Transformation for 3D Gaussians}

The transformation of 3D Gaussians into meshes builds upon the methodology used for flat Gaussians, but extends it to three dimensions. While the 2D approach approximates Gaussians as flat ellipses, the 3D approach represents them as ellipsoids, capturing their full spatial extent. The process is governed by three hyperparameters, $scale\_mul$, $no\_triag$, and $opac\_mul$, which maintain the same values as in the 2D case. However, it should be noted that several significant distinctions emerge due to the augmented dimensionality. Thus, unlike the 2D case, where the smallest scale and its corresponding rotation vector are discarded, all three scale constants and rotation vectors are retained for 3D Gaussians. The transformation process is shown in Fig.~\ref{fig:modle}.

\begin{figure*}[t!] 
  \centering
  \includegraphics[width=0.9\textwidth]{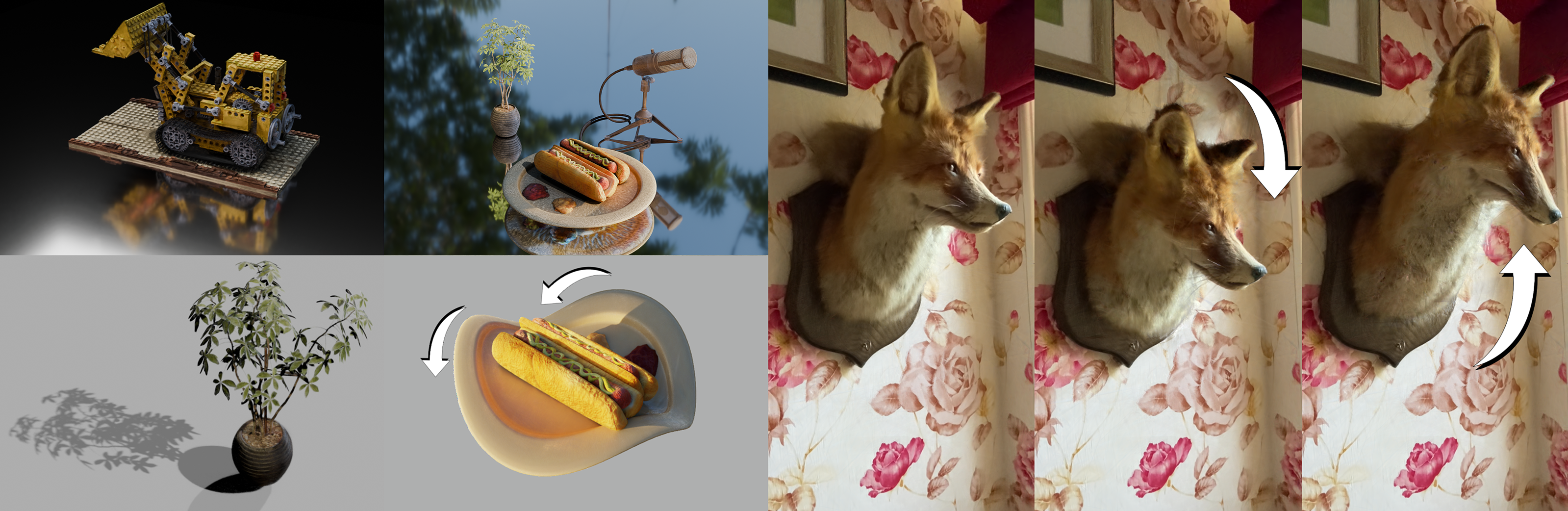}
  \caption{\our{} allows high-quality renders and simulations in classical rendering tools like Blender and Nvdiffrast.}
  \label{fig:editing}
\end{figure*} 

First, three orthogonal vectors defining the principal axes of an ellipsoid are computed as follows:
\begin{equation}\label{eq:3d_scaledrot}
\begin{aligned}
    \text{scaled\_rot}_1 &= scale\_mul \cdot \exp(s_1) \cdot \rr_1, \\
    \text{scaled\_rot}_2 &= scale\_mul \cdot \exp(s_2) \cdot \rr_2, \\
    \text{scaled\_rot}_3 &= scale\_mul \cdot \exp(s_3) \cdot \rr_3,
\end{aligned}
\end{equation}
where $\rr_1$, $\rr_2$, and $\rr_3$ are respective columns of the rotation matrix $R$. These vectors define the orientation and extent of the ellipsoid in 3D space.
Next, $no\_triag$ points are generated at the boundary of each of the three orthogonal surfaces defined by the pairs: 
\begin{equation}
    \begin{array}{c}
(\text{scaled\_rot}_1, \text{scaled\_rot}_2),\\ (\text{scaled\_rot}_1, \text{scaled\_rot}_3), \\(\text{scaled\_rot}_2, \text{scaled\_rot}_3).\\
\end{array}
\end{equation}
For each surface, $no\_triag$ uniformly distributed angles $\theta_i$ in the range $[-\pi,\pi]$ are computed. Then, for every angle $\theta_i$, the vertex $\text{v}_i$ is calculated in a similar way as for 2D method:
\begin{equation}
    \text{v}_i = \m + \cos(\theta_i) \cdot \text{scaled\_rot}_a + \sin(\theta_i) \cdot \text{scaled\_rot}_b,
\end{equation}
where $\text{scaled\_rot}_a$ and $\text{scaled\_rot}_b$ are vectors that span the current surface. 
Shared points at surface intersections avoid redundant vertices and ensure a seamless mesh.

The mesh is then constructed as a collection of triangle fans, assigning one fan to each surface. Each triangular face is defined analogously to the 2D method, resulting in $3 \cdot no\_triag$ triangles per Gaussian. This structure approximates the 3D Gaussian as a set of connected triangles distributed across three orthogonal surfaces. Color and opacity are assigned to each vertex in the mesh in the same manner as in the 2D solution.

This process transforms each 3D Gaussian into a mesh in much the same way as in the previous case. The main difference is the use of three orthogonal faces to capture the entire 3D structure, and the sharing of vertices between faces to maintain mesh efficiency. However, this representation has more vertices and triangles than the 2D case, resulting in a model with higher memory usage and slower rendering.

\paragraph{Rendering \our{} in Nvdiffrast}

The rasterization process, facilitated by Nvdiffrast on a mesh soup, involves transforming 3D vertices into 2D screen space, culminating in rendering these vertices as an image. The process initiates with transforming vertices into clip space utilizing a model-view-projection (MVP) matrix, which maps them into normalized device coordinates. This ensures that the vertices are appropriately positioned for rendering on the screen.

The rasterization process utilizes a depth-peeling technique to manage transparency and overlapping geometry. This method involves the iterative rendering of the scene in multiple layers, ensuring that each layer is correctly composited with the others. The rasterized output and its derivatives are computed for each layer, enabling precise interpolation of vertex colors across the triangles. The interpolated colors are then blended with the accumulated colors from previous layers, taking into account transparency through the alpha channel. This blending process ensures that transparent regions are accurately represented in the final image.

Once all layers have been processed, the final RGB and alpha values are extracted from the accumulated color buffer. The RGB values represent the rendered image, while the alpha channel captures the transparency. This approach takes advantage of Nvdiffrast's efficiency and flexibility to render complex objects, making it particularly suitable for scenes with detailed geometry and transparency. Depth peeling ensures that overlapping elements are handled correctly, producing high-quality renderings.

\paragraph{Fine-tuning of \our{}}

\begin{figure}[t!] 
    \centering
    \makebox[\columnwidth][c]{%
        \begin{tabular*}{\dimexpr\columnwidth-2cm}{@{\extracolsep{\fill}} c c }
            Before optimization & After optimization
        \end{tabular*}
    }\\[1mm] 
    \includegraphics[width=\columnwidth]{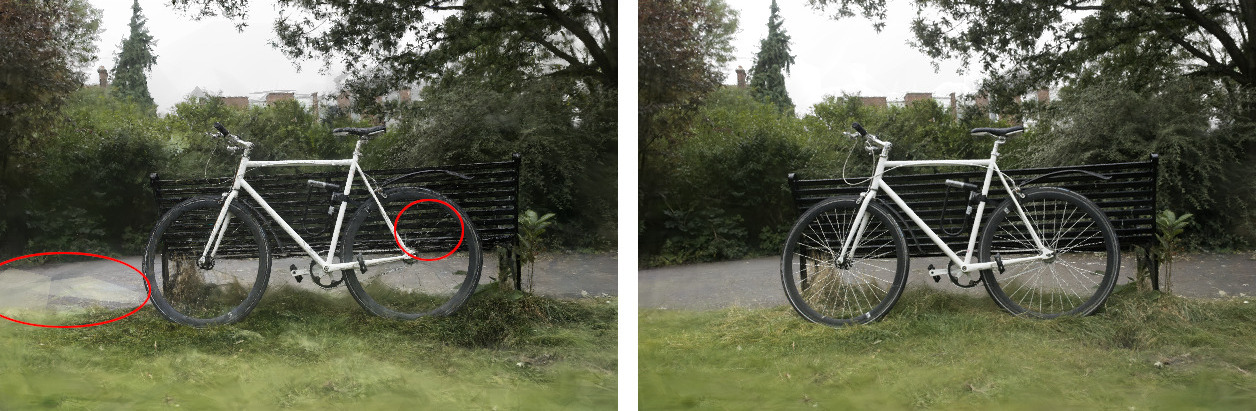}
    \caption{Comparison of the same scene before (left) and after (right) optimization rendered with Nvdiffrast. The unoptimized \our{} rendering is characterized by a lack of fine structural details (red), such as missing or poorly defined bicycle spokes. In contrast, the optimized \our{} outcome shows significant refinement, with the floaters completely removed and finer details clearly reconstructed.
}
    \label{fig:nv_optim_nooptim}
\vspace{-0.4cm}
\end{figure}

Our fine-tuning pipeline accepts the mesh soup created from the Gaussians (either 2D or 3D) as input and refines it to match the target scene better, see Fig.~\ref{fig:nv_optim_nooptim}. The pipeline consists of several stages, including transformation, rasterization, blending, and loss computation, followed by iterative optimization and pruning. Here, we provide a thorough overview of each step in the pipeline.

{\em Transformation of Vertices.} As previously mentioned, the input mesh soup vertices are transformed using the MVP matrix. This matrix projects the 3D vertices into the 2D image space for rasterization. Each vertex $\text{v}_i$ is then transformed as follows: 
$
    \text{v}_{i}^\text{transformed} = \text{MVP}\cdot \text{v}_i.
$

{\em Rasterization and Blending.} The rasterization and blending processes adhere to the methodology above, employing depth peeling to manage overlapping meshes with varying opacities. The output image is derived by integrating multiple rasterized layers, thereby ensuring the preservation of each mesh element's contribution.

{\em Loss Function.} The loss function used for optimization constitutes a weighted combination of $L_1$ loss and the structural similarity index measure (SSIM). This approach is analogous to that utilized in 3DGS, with loss defined as:
\begin{equation}\label{eq:loss}
    \mathcal{L} = \lambda \cdot L_1(\hat{y}, y) + (1 - \lambda) \cdot \text{SSIM}(\hat{y}, y),
\end{equation}
where $y$ is the ground truth image, $\hat{y}$ is the predicted image, and $\lambda = 0.6$ is a balancing constant. 

{\em Optimization and Pruning.} The optimization process updates the mesh soup vertices to minimize the loss function given in Eq.~\eqref{eq:loss}. Specifically, the colors and opacities of the vertices are optimized with a learning rate $lr_\text{color}$. The positions of the vertices are also optimized, but with a weaker learning rate of $lr_\text{verts}=\exp(-3) \cdot lr_\text{color}$, to ensure that the position updates are subtle and stable.

Every 10 epochs, a pruning step is performed to remove redundant faces and vertices. A face is pruned if all its vertices have an opacity below an $\exp(-4)$ threshold. Additionally, if any other face does not use the vertices of the pruned face, they are also deleted. This pruning step helps reduce memory usage and computational complexity while maintaining the fidelity of the scene representation.


\section{\our{} with Mesh Prior}


When using \our{} with a GS prior, we achieve high-quality renderings through a simple and efficient transformation. However, the resulting mesh is optimized solely for rendering and lacks precise geometric accuracy, making it unsuitable for detailed shape approximation. Editing these meshes in Blender resembles manipulating point clouds, akin to the approach in GaMeS~\citep{waczynska2024games}. To overcome this, we integrate external mesh estimation tools as priors~\citep{guedon2023sugar}.

\begin{table*}[t]
\centering
\small
\setlength{\tabcolsep}{4pt}

\caption{Quantitative comparison of \our{} with baseline models on the Mip-NeRF360, Tanks and Temples, and Deep Blending datasets.
On Mip-NeRF360 and Tanks and Temples, \our{} achieves comparable results to rasterization-based techniques despite its mesh-based representation. 
In addition, \our{} shows superior performance on Deep Blending, proving its effectiveness for indoor geometries. 
Note that LinPrim, MobileNeRF, NeRF2Mesh, and GS-IR lack results on the Tanks and Temples and Deep Blending datasets, while RadiantFoam only lacks results on the latter.}
\label{tab:scenefull}
\resizebox{\textwidth}{!}{
\begin{tabular}{llccc ccc ccc}
\toprule
 &  & \multicolumn{3}{c}{Mip-NeRF360} 
    & \multicolumn{3}{c}{Tanks and Temples} 
    & \multicolumn{3}{c}{Deep Blending} \\
\cmidrule(lr){3-5} \cmidrule(lr){6-8} \cmidrule(lr){9-11}
 &  & SSIM $\uparrow$ & PSNR $\uparrow$ & LPIPS $\downarrow$
    & SSIM $\uparrow$ & PSNR $\uparrow$ & LPIPS $\downarrow$
    & SSIM $\uparrow$ & PSNR $\uparrow$ & LPIPS $\downarrow$ \\
\midrule

\multirow{11}{*}{\rotatebox{90}{Spherical Harmonics}}
& MobileNeRF   & 0.527 & 23.06 & 0.430 & -     & -     & -     & -     & -     & - \\
& NeRF2Mesh    & 0.523 & 22.74 & 0.460 & -     & -     & -     & -     & -     & - \\
& Plenoxels    & 0.670 & 23.63 & 0.440 & 0.379 & 21.08 & 0.795 & 0.510 & 23.06 & 0.510 \\
& INGP-Base    & 0.725 & 26.43 & -     & 0.723 & 21.72 & 0.330 & 0.797 & 23.62 & 0.423 \\
& INGP-Big     & 0.751 & 26.75 & 0.300 & 0.745 & 21.92 & 0.305 & 0.817 & 24.96 & 0.390 \\
& M-NeRF360    & 0.844 & \textbf{29.23} & -     & 0.759 & 22.22 & 0.257 & 0.901 & 29.40 & 0.245 \\
& 3DGS-30K     & \textbf{0.87} & 28.69 & 0.220 & \textbf{0.841} & 23.14 & \textbf{0.183} & \textbf{0.903} & \textbf{29.41} & \textbf{0.243} \\
& 3DGRT        & 0.854 & 28.71 & 0.250 & 0.830 & \textbf{23.20} & 0.222 & 0.900 & 29.23 & 0.315 \\
& GS-IR        & 0.812 & 26.57 & 0.238 & -     & -     & -     & -     & -     & - \\
& LinPrim      & 0.803 & 26.63 & 0.221 & -     & -     & -     & -     & -     & - \\
& RadiantFoam  & 0.830 & 28.47 & \textbf{0.210} & - & - & - & 0.890 & 28.95 & 0.260 \\
\midrule

\multirow{2}{*}{\rotatebox{90}{RGB}}
& RaySplats    & \textbf{0.846} & 27.31 & 0.237 & \textbf{0.829} & \textbf{22.20} & \textbf{0.202} & \textbf{0.900} & \textbf{29.57} & 0.320 \\
& \our{} (our) & 0.817 & \textbf{28.08} & \textbf{0.229} & 0.766 & 21.71 & 0.248 & 0.890 & 29.50 & \textbf{0.254} \\

\bottomrule
\end{tabular}
}
\end{table*}

In the \our{} framework with a mesh prior, we leverage a geometry mesh---estimated using external tools---that accurately represents the shape of the object. On top of this geometry mesh, \our{} constructs a secondary layer composed of disjoint, colored triangles with associated opacity attributes. This approach, which utilizes triangular elements instead of the fan-type meshes typical for \our{} with a GS prior, significantly reduces the total face count and enables precise control over both rendering quality and manual editing. Training is efficiently constrained to the surface of the geometry mesh, and the initialization of the secondary triangles is directly based on the faces of this mesh.
The parameterization of these triangles is both flexible and consistent. Consider a single triangular face of the geometry mesh with vertices:  
$
M = \{ \mathbf{m}_1, \mathbf{m}_2, \mathbf{m}_3 \} \subset \mathbb{R}^3.
$  
Each vertex $\mathbf{v}_i$ of a new triangle $V = \{\mathbf{v}_1, \mathbf{v}_2, \mathbf{v}_3\}$ is expressed as a convex combination of the geometry face vertices:  
$
\mathbf{v}_i(\alpha_1, \alpha_2, \alpha_3) = \alpha_1 \mathbf{m}_1 + \alpha_2 \mathbf{m}_2 + \alpha_3 \mathbf{m}_3,
$  
where the trainable parameters $\alpha_1, \alpha_2, \alpha_3$ satisfy $\alpha_1 + \alpha_2 + \alpha_3 = 1$ and $\alpha_1, \alpha_2, \alpha_3 \geq 0$, see Fig. \ref{fig:mesh_modification}. This ensures that all generated triangles remain within the bounds of the original geometry mesh and, therefore, are optimally positioned for rendering.

\begin{figure*}[t]
\centering
\includegraphics[width=1.0\textwidth]{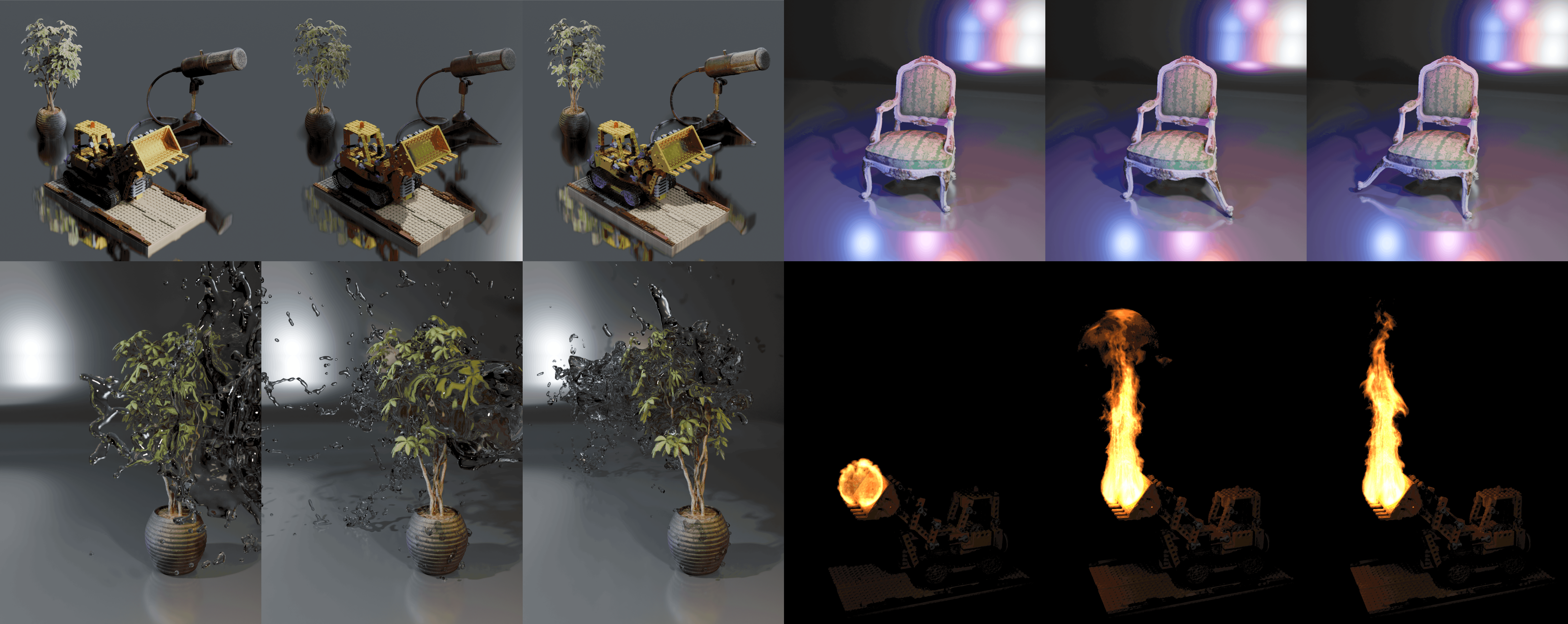}
\caption{Renders obtained by \our{} in  Blender. Top-left shows working reflections with changing position of the light source. Bottom-left - physical interaction between \our{} asset and water. Top-right effect of editing structure of \our{} asset. Bottom-right, effect of expanding source of light on \our{} asset.
}
    \label{fig:additional_1}
\vspace{-0.3cm}
\end{figure*}

To maintain consistency and efficiency, \our{} employs a fixed number of triangles per mesh face. The entire model is fine-tuned using Nvdiffrast, following a process analogous to that used for \our{} with a GS prior. This hierarchical and parameterized structure facilitates seamless integration with standard graphics pipelines and manual mesh editing workflows allowing advanced editing possibilities, see Fig. \ref{fig:additional_1}.

\section{Experiments}

In this section, we present the outcomes of our experimental study. For all quantitative comparisons (see Tab.~\ref{tab:scenefull} and Tab.~\ref{tab:mip_nerf360_results}), we employed \our{} with a GS prior for optimal visual quality, as our mesh prior mode prioritizes editability and geometric control.
Furthermore, we elucidate the behavior of our algorithm after parametrization and training. 

\paragraph{Quantitative Results}

We evaluated the performance of \our{} across three benchmark datasets.
The quantitative outcomes are summarized in Tab.~\ref{tab:scenefull}. On the Mip-NeRF360 and Tanks and Temples datasets, results were observed to be comparable to existing methods, with SSIM, PSNR, and LPIPS metrics aligning closely with state-of-the-art approaches such as 3DGS~\citep{kerbl20233d} and RaySplats~\citep{byrski2025raysplats}. For instance, on the Mip-NeRF360 dataset, our method achieved scores comparable to rasterization-based techniques despite its mesh-based representation. Similarly, on the Tanks and Temples dataset, performance metrics remained competitive, reflecting robustness in outdoor scene reconstruction. It is noteworthy that \our{} exhibited superior performance on the Deep Blending dataset. This outcome suggests that our method is effective in managing intricate indoor geometries and transparency effects, surpassing several baseline methods and approaching the performance metrics of models such as 3DGS.

\paragraph{Qualitative Results}

\our{} renders obtained using Nvdiffrast exhibited a visual quality nearly identical to 3DGS without spherical harmonics, as illustrated in Fig.~\ref{fig:big-scene_nvdiffrast}. Both methods demonstrated similar limitations in modeling reflections on transparent surfaces, such as glass, where approximations of light interactions led to artifacts. Slight reductions in detail were detected in high-frequency regions, such as dense grass, where \our{} supplied marginally coarser reconstructions. Notwithstanding these minor discrepancies, the overall fidelity of the outcomes of our model was found to be comparable to those produced by 3DGS, thereby underscoring its ability to replicate the strengths of Gaussian-based rendering within a mesh-based framework.


Furthermore, when rendered using Blender's EEVEE engine, \our{} also produced visually plausible results comparable to 3DGS (see Fig.~\ref{fig:editing}). However, challenges persisted in areas requiring intricate geometric precision, where high-frequency details were partially lost—nevertheless, the majority of scenes retained photorealistic quality, with coherent geometry and accurate color representation. 

\paragraph{SuGaR with UV Texture}

Generating meshes from 3DGS representations is also critical for editable 3D assets, yet existing methods like SuGaR \citep{guedon2023sugar} produce geometry meshes without inherent color attributes, instead relying on UV textures for rendering---a decoupled approach that underperforms surface---aligned 3DGS in metrics like PSNR, SSIM, and LPIPS \citep{huang20242d,szymkowiak2024neural}. In contrast, \our{} directly converts trained 3DGS into a colored mesh by parameterizing Gaussians via the mesh structure, eliminating texture training and outperforming both UV-textured meshes and surface-aligned 3DGS in rendering quality, as shown in Tab.~\ref{tab:mip_nerf360_results}. While SuGaR prioritizes editability at the cost of fidelity, \our{} delivers high-quality renders while retaining compatibility with standard editing tools.


\begin{table}[t!] 
\caption{Quantitative comparison of \our{} with surface-aligned 3DGS
and an optimized traditional UV texture on the Mip-NeRF360 dataset. }
\centering
\label{tab:mip_nerf360_results}
{ 
\begin{tabular}{@{\;\;}c@{\;\;\;\;}c@{\;\;}c@{\;\;}c@{\;\;}c@{}}
\toprule
Method & PSNR $\uparrow$ & SSIM $\uparrow$& LPIPS $\downarrow$ \\
\midrule
1M vertices (3DGS) & 24.51 & 0.768 & 0.295 \\
1M vertices (UV) & 21.24 & 0.609 & 0.478 \\
\midrule
200K vertices (3DGS) & 24.24 & 0.757 & 0.300\\
200K vertices (UV) & 21.44 & 0.656 & 0.419 \\
\midrule
MeshSplats & 28.08 & 0.817 & 0.229\\
\bottomrule
\end{tabular}
}
\end{table}

\section{Conclusions}

In this paper, we present \our{}, a method that addresses the limitations of GS by transforming it into a mesh-like structure compatible with ray tracing. This enables enhanced rendering with improved lighting, shadows, and reflections. \our{} provides an efficient and practical solution that can be further refined through our optimization algorithm. Extensive experiments confirm its effectiveness and versatility across diverse datasets. Despite using a mesh-based representation, \our{} achieves photorealistic quality that is comparable to or better than GS, eliminating floaters and refining structural details. These results highlight \our{} as a strong alternative for high-quality rendering in computer graphics.

\paragraph{\bf Limitations}
The main limitation of \our{} is that artifacts such as fragmented geometry and inconsistent opacity can appear in large, low-texture areas (e.g., the sky in the Train scene), due to the mesh-based interpolation struggling to mimic the smooth fall-off of Gaussians.


\let\clearpage\relax
\vspace{1.5\baselineskip}

\bibliographystyle{unsrt}

\end{document}